\documentclass[]{spie}  
\usepackage{amsmath,amsfonts,amssymb}
\usepackage{graphicx}

\title{On-Sky Operations with the ALES Integral Field Spectrograph}

\author[a]{Jordan M. Stone}
\author[b]{Andrew J. Skemer}
\author[a]{Philip Hinz}
\author[b]{Zack Briesemeister}
\author[c]{Travis Barman}
\author[d]{Charles E. Woodward}
\author[e]{Mike Skrutskie}
\author[a]{Jarron Leisenring}
\affil[a]{Steward Observatory, University of Arizona, 933 N. Cherry Ave., Tucson, AZ, 85721, USA}
\affil[b]{University of California, Santa Cruz,1156 High Street, Santa Cruz, CA, 95064, USA}
\affil[a]{University of Arizona, 933 N. Cherry Ave., Tucson, AZ, 85721, USA}
\affil[c]{Lunar and Planetary Laboratory, The University of Arizona, 1629 E. Univ. Blvd., Tucson, AZ 85721 USA}
\affil[d]{Minnesota Institute of Astrophysics, University of Minnesota, 116 Church Street, SE, Minneapolis, MN 55455, USA}
\affil[e]{Department of Astronomy, University of Virginia, Charlottesville, VA 22904, USA}

\begin{document} 
\maketitle

\begin{abstract} The integral field spectrograph configuration of the LMIRCam science camera within the Large Binocular Telescope Interferometer (LBTI) facilitates 2 to 5~$\mu$m spectroscopy of directly imaged gas-giant exoplanets. The mode, dubbed ALES, comprises magnification optics, a lenslet array, and direct-vision prisms, all of which are included within filter wheels in LMIRCam. Our observing approach includes manual adjustments to filter wheel positions to optimize alignment, on/off nodding to track sky-background variations, and wavelength calibration using narrow band filters in series with ALES optics.  For planets with separations outside our 1''x1'' field of view, we use a three-point nod pattern to visit the primary, secondary and sky. To minimize overheads we select the longest exposure times and nod periods given observing conditions, especially sky brightness and variability. Using this strategy we collected several datasets of low-mass companions to nearby stars.  \end{abstract}


\section{INTRODUCTION} \label{sec:intro}  

At low temperatures gas-giant atmospheres transition from relatively warm,
cloudy, and CO-rich (L-types) to relatively cool, and cloud free with CH$_4$
not CO as the dominant carrier of atmospheric carbon (T-types). This transition
is not completely understood especially because it is so abrupt and dramatic.
For example, the field brown dwarf population transition beginning at
$T_{\mathrm{eff}}$$\sim1300$~K and ending by $T_{\mathrm{eff}}\sim1100$~K
(e.g., Ref. \citenum{Kirkpatrick2005}). 

The temperature of the L-to-T transition is a function of gravity, with
low-gravity objects observed to loft photospheric clouds and maintain
atmospheric CO down to much lower temperatures than their higher-gravity
counterparts (e.g., Ref. \citenum{Stephens2009}).  Thus, degeneracies between
surface gravity, effective temperature, cloudiness, and chemistry make
interpreting spectra of substellar objects challenging.  Thermal-infrared
(3--5$\mu$m) measurements are essential for breaking these degeneracies because
they probe the peak of the emission spectra for cool
($T_{\mathrm{eff}}\lesssim1000$~K) atmospheres and because they probe the
fundamental transitions of the CH$_{4}$ and CO molecules. Thus, 3--5~$\mu$m
spectral energy distributions are sensitive indicators for understanding carbon
chemistry in substellar objects\cite{Noll2000}.

While essential for a complete understanding of planetary atmospheres,
thermal-infrared characterization of exoplanets has received relatively little
attention, partly because of the challenging nature of ground-based
observations at these wavelengths. The LBTI instrument at the LBT  was
specifically designed to provide low thermal background for the study of
exoplanetary systems~\cite{Hinz2016}. The instrument takes advantage of the
observatory's twin deformable mirror adaptive optics systems to observe at
very-high spatial resolution using the minimum number of warm optics. 

To facilitate thermal-infrared spectroscopic observations of exoplanets, our
team has built the world's only thermal infrared integral field
spectrograph~\cite{Skemer2015}. Integral field spectrographs (IFS) facilitate
exoplanet spectroscopy because the techniques of high-contrast imaging can be
used to separate planet light from starlight at each wavelength. The
instrument, named the Arizona Lenslets for Exoplanet Spectroscopy (ALES), has
recently been commissioned and is now producing high-spatial resolution L-band
spectra in high-contrast environments.

The ALES IFS was built into the existing LMIRCam module~\cite{Skrutskie2010}
of LBTI. To accomplish this, all the ALES optics, including magnifiers,
a lenslet array, and direct-vision prisms, are located within LMIRCam filter
wheels. This architecture requires a careful set-up and calibration approach
that accommodates the number of moving parts. 

In this manuscript, we document our ALES observing strategy which we developed
over three semesters while using ALES to observe multiple systems including
exoplanets, brown dwarfs and solar system objects.  This summer, ALES will
undergo a significant upgrade that will improve our usable field of view,
spectral resolution, and overall optical quality. New Magnification optics will
also make imaging interferometry possible with ALES. (See Hinz et al., and
Skemer et al., in these proceedings for more specifics.) After the upgrade,
some of the details of our observing strategy may change, but the
overall process and general steps of our approach will remain unchanged. 

\section{Aligning Optics} 

\begin{figure} [ht]
\begin{center}
\includegraphics[angle=90, height=17cm]{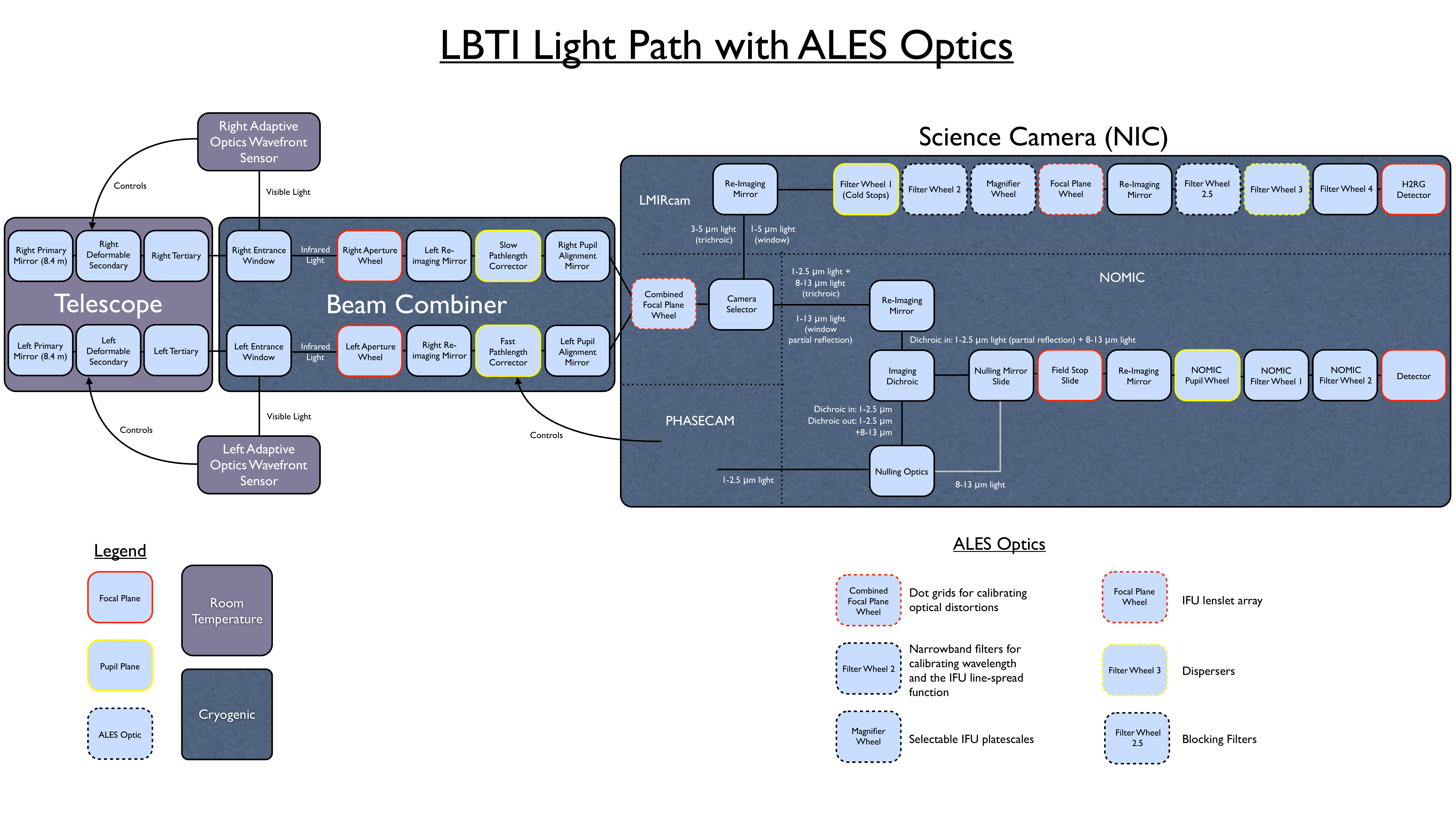}
\end{center}
\caption{Block diagram of LBTI showing the full ALES light path. All of the
ALES optics are inserted by filter wheels, making the design completely
modular. We use the NOMIC science camera in parallel with ALES when we
implement a three-point nod script to observe widely separated companions. Not
shown: four flat mirrors, a ZnSe window between the beam combiner and the
science camera, and a fixed pupil stop in the NOMIC light path.
 \label{fig:blocks}}
\end{figure} 

In the following subsections we will refer to elements indicated in Figure
\ref{fig:blocks} in order to describe the set-up and alignment of ALES. In some
cases the elements are named different than indicated within the LBTI control
software actually used to set up the instrument. In these cases, we list the
alternative name in parentheses.

The unique architecture of the LBT telescope, with its two 8.4 diameter mirrors
separated by 6.2 meters, facilitates multiple observing modes including
co-phased imaging interferometry as well as separated imaging where two images
of a target star are made on the detector. In its preliminary configuration
ALES is used in single-aperture mode with LBTI as the current
magnification optics do not deliver an ALES plate scale that can Nyquist sample
the interferometric point spread function (34 milliarcseconds at 3.8 $\mu$m).
The ALES field of view is too narrow to make use of dual-sided
but spatially separated imaging.  An alternative option, overlapped but not
co-phased imaging with LBTI, is possible with ALES in principle, but in
practice, added operational overheads associated with overlapping the two sides
make this mode sub-optimal. A planned ALES upgrade will include new
magnification optics that will allow ALES to be used in imaging interferometry
mode (Skemer et al., these proceedings).

Single-sided imaging with LBTI/LMIRCam is typically done using a combination of
the dual-sided cold stop in LMIRCam Filter Wheel 1 (lmir FW1) and an opaque
semi-circle ``half moon" in LMIRCam Filter Wheel 2 (lmir FW2). The ``half moon"
stops the light from the unused side of the telescope in order to reduce the
background on the detector. This approach is preferred for most programs
requiring only one side with LMIRCam because it is fast to move the ``half
moon" in and out when switching between programs with different instrument
requirements.  However, the ALES wavelength calibration filters are also
located in Filter Wheel 2 (lmir FW2), so the dual-sided cold stop plus ``half
moon" approach cannot be used with ALES. Instead, we use a single-sided cold
stop in Filter Wheel 1 (lmir FW1).  This adds $\sim5$ mins to the set-up
overheads for ALES compared to other single-sided modes, and another $\sim5$ mins at
the end of an ALES observing sequence to replace the dual-sided cold stop for
use with other programs.

\subsection{Align the Single-Sided Cold Stop}
The process for aligning the single-sided cold stop begins by entering
pupil-imaging mode. This entails:
\begin{itemize}
\item Camera Selector (NIC BEAMDIVT) goes to Trichroic,
\item Combined Focal Plane Wheel (NIL NICNAC) goes to 5mm pinhole,
\item Filter Wheel 2 (lmir FW2) goes to Open,
\item Filter Wheel 2.5 (lmir FW25) goes to Open,
\item Filter Wheel 3 (lmir FW3) goes to narrowband 3.9 $\mu$m filter,
\item Filter Wheel 4 (lmir FW4) goes to pupil imaging lens.
\end{itemize}
The next step is to decide which side of the telescope to use. Typically which
side is chosen is unimportant, but during intervals of telescope technical
difficulties it may be that one side is better suited to ensure stable system
performance for science operations. Then the corresponding cold stop, either
single-sided left (SXAperture) or single-sided right (DXAperture) should be
selected for Filter Wheel 1 (lmir FW1). The position of the stop should be
optimized to limit thermal background. Gross vertical alignment is made using
small adjustments (~1000 steps) to Filter Wheel 1 (lmir FW1).  Fine adjustments
and horizontal adjustments are made using the corresponding right or left side
Pupil Alignment Mirrors (roof mirrors). The configured maximum single movement
for the roof mirrors is 10,000 steps, and this results in a relatively small
    motion of the pupil. 

\subsection{Align the Target with ALES Sweetspot}
Once the cold stop is aligned, LMIRCam should be returned to L$^{\prime}$
imaging mode:
\begin{itemize}
\item Combined Focal Plane Wheel (NIL NICNAC) goes to Open,
\item Filter Wheel 3 (lmir FW3) goes to Open,
\item Filter Wheel 4 (lmir FW4) goes to std-L.
\end{itemize}

The next step is to choose a bright star to facilitate ALES alignment. Stars
with m$_{\mathrm{L}^{\prime}}\lesssim7$ will allow all alignment steps to be
carried out with minimum exposure times. Typically a telluric calibrator or
nearby pointing star is used if the science object is fainter than this.

In imaging mode, the image of the target needs to be offset to be coincident
with the ALES sweetspot---the portion of the LMIRCam field-of-view that is seen
when the ALES magnifier is used. In its current set up the sweet spot is
centered on LMIRCam pixel (1388,996). Before offsets can be made, the
adaptive-optics loop for the relevant side needs to be locked. After verifying
that the AO loop is closed, the required offset in detector coordinates can be
calculated using the LMIRCam plate scale, 10.07 milliarcseconds/pixel\cite{Maire2015}. 

The current version of ALES (before an upgrade this summer), suffers from
off-axis astigmatism. The best image quality is centered around pixel (900,
1040) in the magnified field of view.  To optimize the position of the target,
      we put in the magnifier and make small telescope offsets, keeping in mind
that the magnified field is 180$^{\circ}$ rotated with respect to the
unmagnified image because the refractive magnifier sends the light through
focus~\cite{Skemer2015}. To make this step easier, it is helpful to first save
a dark image and then subtract it from subsequent frames before displaying on
the graphical interface. The process is:

\begin{itemize}
\item Filter Wheel 4 (lmir FW4) goes to Blank,
\item take an image and use it as a background on the display,
\item Filter Wheel 4 (lmir FW4) goes to std-L,
\item Magnifier Wheel (Mag Wheel) goes to Refractive Magnifier,
\item make small adjustments to the target position with telescope offsets,
keeping in mind the field is rotated.
\end{itemize}


Once the star is positioned as desired, the AO config file should
be updated with the corresponding bayside stage positions so that new pointings
are acquired with the star at the same location. The offsets should also be
``absorbed" by the telescope Pointing Control System so that an absolute (0, 0)
offset will return the star to it's optimal position.

\subsection{Align ALES Optics}

The ALES configuration includes a magnifier, a lenslet array, a prism and
a blocking filter. After following the steps above, the magnifier should
already be in place in the Mag Wheel. To complete the set-up we align the
lenslet array in the Focal Plane Wheel (lmir APERWHL) and the prism in Filter
Wheel 3 (lmir FW3).

First, select the lenslet array in the Focal Plane Wheel (lmir APERWHL). It is
important to verify that the lenslet array is properly positioned, meaning that
the lenslet spots are parallel to the pixels on the LMIRCam detector. Using
a 2-second exposure time is helpful for this step and grabbing a corresponding
dark frame for background subtraction in the graphical interface is also
helpful. Currently the displayed image in our graphical interface is binned to
$512\times512$ pixels. This can make analyzing the lenslet spots challenging.
Our solution is to zoom in on the $512\times512$ pixel region including the
highest optical quality spots. The correct way to zoom is done by selecting the
appropriate subframe for the detector and using LMIRCam ``slow mode". In this
context, slow-mode does not imply any change in the detector read-out. Rather,
the whole $2048\times2048$ detector is read and saved but only the pixels
within the chosen subframe are displayed.  This is peculiar to our graphical
interface implementation, and allows us to save full ALES frames while viewing
full resolution images of a smaller subframe. Once zoomed in, small nudges to
the Focal Plane Wheel (lmir APERWHL) can be used to adjust the lenslet array
for optical positioning.

Next, the prism and blocking filter should be positioned. Currently ALES has
only one spectral mode covering 2.8 through 4.2 $\mu$m at R$\sim40$. In the
future, a wider variety of passbands and spectral resolutions will be available
(Skemer et al., these proceedings). Prisms are in Filter Wheel 3 (lmir FW3),
and blocking filters are in Filter Wheel 2.5 (lmir FW25).  Once positioned the
dispersion direction of the spectral traces should be inspected. The optimal
dispersion direction results in minimal overlap of neighboring spectra. If
spectra appear to collide, the best recourse is to home Filter Wheel 3 (lmir
FW3) and then re-select the prism.

In list form the above sequence is:
\begin{itemize}
\item Focal Plane Wheel (lmir APERWHL) goes to lenslet array,
\item choose $512\times512$ subframe and set to slow mode,
\item Filter Wheel 4 (lmir FW4) goes to Blank,
\item save a dark frame for background subtraction,
\item Filter Wheel 4 (lmir FW4) goes to Open,
\item adjust lenslet array position if needed,
\item Filter Wheel 3 (lmir FW3) goes to IFU-prism,
\item Filter Wheel 2.5 (lmir FW25) goes to Lspec,
\item home and re-position Filter Wheel 3 (lmir FW3) if needed.
\end{itemize}

Once aligned, it is very important not to touch any of the ALES-specific optics
(magnifier, lenslet, prism) before obtaining wavelength calibration frames
because the wavelength solution depends on the specific position of all of
these elements.

\section{Observing} 
Since LBT is and Alt/Az telescope and the LBTI instrument does not include
a derotator, all objects appear to rotate with the parallactic angle on the
LMIRCam detector. In closed-loop adaptive optics mode, the AO-guidestar is the
axis of rotation. This paradigm requires two ALES observing scenarios,
depending on whether or not the object of interest can be seen within the
narrow ALES field of view together with the guide star.

\subsection{Two-point Nod Sequence} For self-guided targets, including telluric
calibrators and solar-system objects (e.g., comets and planetary moons), and
for close-in directly imaged exoplanets such as HR~8799~c,d,e (with the primary
as the AO-guidestar), we observe with a two-point nod script periodically
pointing to the sky in order to track variable background emission.  Exposure
times for the camera are chosen to be as long as possible without detector
saturation. For close-in planets this requirement is relaxed so that pixels
covering the core of the stellar PSF are often saturated, though it is
important to ensure that pixels at the separation of interest are still linear.
With the current LMIRCam set-up, significant non-linearity occurs above 50,000
counts (i.e., ADUs). Typically exposure times are 2 to 3 seconds long. The
number of images per nod position is selected so that the telescope is offset
every 2 mins.

In order to track and flag highly variable pixels that complicate the
extraction and analysis of ALES data, we collect pseudo-dark frames
simultaneously with telescope nods. To do this, we spin Filter Wheel 2 (lmir
FW2) by 25,000 steps (half a position) and save two or three frames (depending
on the exposure time) while the telescope is in motion, thus there are no
additional overheads associated with saving these frames.

Exposing science and sky-background frames, nodding, and saving darks while the
telescope is in motion are all performed within a script for efficiency. The
script coordinates with the adaptive-optics system and begins collecting
science and sky data only after AO loops are closed following a telescope move.
To facilitate the organization and analysis of ALES data, the script
automatically updates the fits headers of each file, setting the FLAG keyword
to SCI, DRK, or SKY, for on-source, pseudo-dark, and sky-background frames,
respectively.

\subsection{Three-point Nod Sequence}
Our alternate observing scenario applies to targets more widely separated from
their guidestar. This is the case for HR~8799~b, $\kappa$~And~b, GJ~504~b, and
HD~130948 BC for example. In these cases we must nod the telescope to track
rotation of the parallactic angle to ensure that our target of interest remains
in the ALES field of view. To achieve this we use a three-point nodding script
that alternates pointing at the guide-star, the science target, and a blank sky
position to track variable background. Exposure times are selected as above
with the added constraint that they should be short enough that the science
target does not smear due to rotation within an integration. This typically
applies only to the widest separation most quickly rotating objects. We spend
two minutes on the science target and two minutes in the sky position for each
nod cycle. Since the guide-star is observed only to help anchor our coordinate
system we save far fewer frames in this position, usually 4. As in the
two-point scenario, our script saves pseudo-darks while the telescope is in
motion and automatically updates fits headers for organization. In this case
the FLAG keyword is set to PRI, SEC, SKY, or DRK for the guide-star, science
target, sky-background, and pseudo-dark frames, respectively.

For our three-point nod script, we use RA and Dec-coordinate absolute (not
relative) nods within the script in order to offset to the science position.
We use these nod parameters because in this system the guidestar-science object
separation vector ($\Delta$RA, $\Delta$Dec) is not a function of time (unlike
detector coordinate or relative offsets which change as the field rotates).  In
this mode, it is particularly important to make sure that any adjustments to
the position of the AO-guidestar are ''absorbed" by the telescope pointing
control system before executing the script.

\begin{figure} [ht]
\begin{center}
\includegraphics[height=12cm]{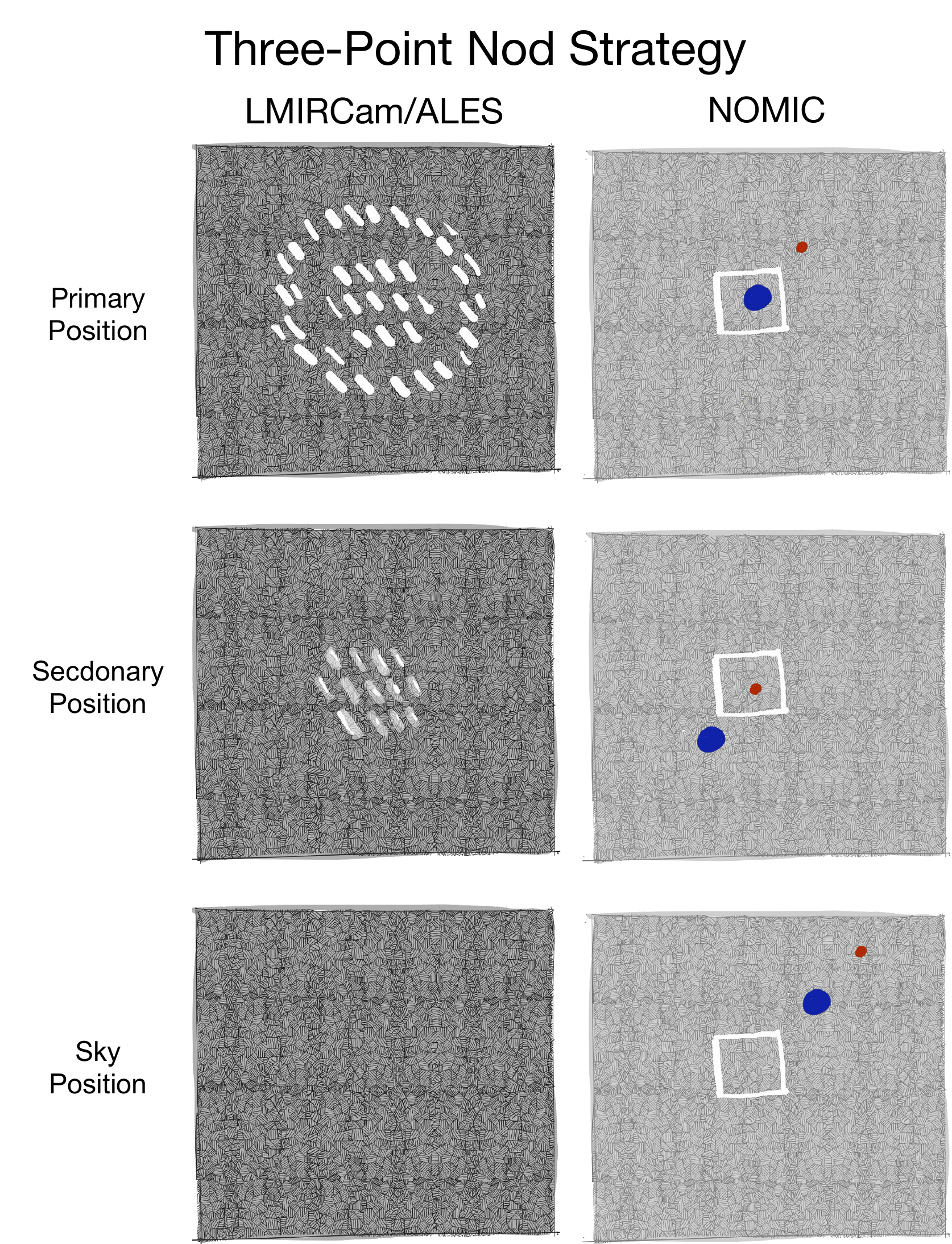}
\end{center}
\caption{A cartoon illustration of our three-point nod script approach to
observing widely separated companions with ALES. The left column represents the
sky-subtracted images seen by ALES. The right column represents the field as
seen by NOMIC. Using the NOMIC frames we can empirically measure the LBT nods,
and use this information to shift and stack faint secondary frames when the
secondary is too faint to see in an individual 2-minute nod.
 \label{fig:3nod}}
\end{figure} 

\subsubsection{Using NOMIC to Track Nods} 

We found that imprecise nods prohibited the naive stacking of faint science
target data collected with our three-point approach. To circumvent this
limitation, we use the NOMIC camera~\cite{Hoffmann2014} which is positioned
within LBTI on the opposite side of a trichroic beam splitter from ALES.  The
beam splitter sends 1--2.5~$\mu$m and 8--13~$\mu$m light toward NOMIC and
3 to 5~$\mu$m light toward LMIRCam (and ALES). Usually, NOMIC is used as
  a 10~$\mu$m nulling interferometer, but for our purposes we use it in H-band
(1.65~$\mu$m) imaging mode to observe our fields in parallel with ALES. In this
way, we can precisely measure the telescope nod-vector using the image of the
guidestar on NOMIC, and then use the information to solve for the appropriate
image shifts needed to stack our faint-star data.  This requires some
additional set-up time for three-point nod observations to get NOMIC ready.

First, NOMIC should be put into imaging mode:
\begin{itemize}
\item Imaging Dichroic (NIL DICROIC) goes to Imaging,
\item Nulling Mirror Slide (NOMIC MSSLIDE) goes to Imaging.
\end{itemize}
Next, NOMIC should be put into pupil imaging mode:
\begin{itemize}
\item NOMIC Filter Wheel 1 (NOMIC FW1) goes to Pupil-imaging-lens,
\item NOMIC Filter Wheel 2 (NOMIC FW2) goes to broad N-band (W-10145-9),
\item Combined Focal Plane Wheel (NIL NICNAC) goes to 5mm pinhole,
\item NOMIC Detector goes to low gain.
\end{itemize}
Since the Pupil Alignment Mirrors (roof mirrors) are positioned to align the
LMIRCam cold stop, we will align the NOMIC cold stop as best we can without
touching them: 
\begin{itemize}
\item NOMIC Pupil Wheel (NOMIC PW) goes to Dual2.54, a double-sided cold stop,
\item small adjustments to NOMIC Pupil Wheel (NOMIC PW) move the stop left-right.
\end{itemize}
NOMIC does not have a single-sided stop like LMIRCam, so to stop light from the
side not in use, we send the appropriate Aperture Wheel in the beam combiner to
Closed. Lastly, we put NOMIC into H-band imaging mode:
\begin{itemize}
\item NOMIC Filter Wheel 1 (NOMIC FW1) goes to Open,
\item NOMIC Filter Wheel 2 (NOMIC FW2) goes to H-band (W0164-6),
\item Combined Focal Plane Wheel (NIL NICNAC) goes to Open,
\item NOMIC Detector goes to High gain.
\end{itemize}

We adjust the NOMIC exposure time to make sure the guidestar does not saturate,
and within our script we set the number of images per nod position such that
NOMIC always takes less time gathering images than ALES.

\section{Calibrations}

To properly interpret ALES images some ALES-specific calibration data is
required. These include wavelength calibration frames, lenslet flats, and
telluric calibrators. 

Wavelength calibration is currently done using a set of four narrow-band
photometric filters in series with the ALES optics, upstream of the lenslet
array. These R$\sim100$ filters are unresolved by the ALES prisms and provide
single-wavelength spots on the LMIRCam detector. We have have cryogenic traces
of our filters and the central wavelengths from short to long are:
2.9, 3.3, 3.5, and 3.9~$\mu$m. For each ALES trace, a second-order polynomial
  is fit to the position of the four spots to derive the wavelength solution.
This approach allows us to identify changes in the wavelength solution as
a function of position as the optical quality degrades off axis.  Typically
$\gtrsim100$ seconds of exposure time is needed for high S/N in the 2.9 $\mu$m
filter, much less is necessary for the 3.9 $\mu$m filter because the background
increases dramatically between these two wavelengths.

\begin{figure} [ht]
\begin{center}
\includegraphics[height=7.5cm]{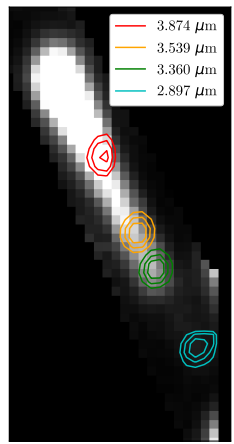}
\end{center}
\caption{ \label{fig:wavecal} 
ALES wavelength calibration strategy. Four unresolved narrow-band filters are
used in parallel with ALES optics to measure the wavelength solution. A single
ALES trace is shown in gray scale. Contours of the narrow-band+ALES images are
overlaid.}
\end{figure} 

Since ALES is composed of so many moving parts, it is essential to collect
wavelength calibration data before moving any of the ALES optics (to move on to
a non-ALES observing program for example). If necessary in a time crunch,
observations through the 3.9 $mu$m filter can be made on-sky and then during
closed-dome time in the morning a full four-filter set can be collected. In
this case, we solve for the rotation and shift necessary to overlap the dome
3.9 $\mu$m frames on top of the sky 3.9 $\mu$m frames. We then apply these
  corrections to the frames from all four filters before performing the
wavelength calibration. This approach is not perfect, since slight
mis-alignments of the lenslet array will put a given lens at a different
location in the intermediate focal plane, where aberrations are a function of
position.

In the future we will have a dedicated wavelength calibration unit for ALES
that will include a light-source and monochrometer (Briesemeister et al., these
proceedings). In addition to helping wavelength calibrate our data, this unit
will facilitate a new spectral extraction mode that will increase our signal to
noise and decrease our susceptibility to spaxel crosstalk~\cite{Brandt2017}.

A lenslet flat is a flat-field that quantifies the through put of each lenslet
in our lenslet array. We do not typically spend time specifically to obtain
data for a lenslet flat. Instead, we use the sky-images observed while
collecting science data since the background is typically bright enough and
very flat across our 1'' field of view.

Telluric calibrations for ALES are similar to telluric calibrations for other
spectrographs. We typically choose to observe A-type stars to calibrate our 2.8
to 4.2 $\mu$m spectra. In the future, when our Brackett-$\gamma$ mode is
implemented, calibrators without strong photospheric hydrogen lines, such as
G-type stars, should be selected. In order to not saturate the LMIRCam
detector, a calibrator fainter than m$_{L^{\prime}}\sim4$ should be chosen. For
efficiency a calibrator brighter then m$_{L^{\prime}}\sim7$ should be
identified. One important and unique requirement for ALES telluric calibration
is to make sure that the calibrator overlaps similar spaxels as the science
object because the spectral resolution changes with position. Howerver, it is not
necessary to get this precise as the resolution is very-low in general and
it changes very slowly across the field. For science targets too faint to see
in individual frames, it is important to step the calibrator about in the ALES
field to ensure that appropriate data is collected.

\section{Initial Results} 
\begin{figure} [ht]
\begin{center}
\includegraphics[height=7.5cm]{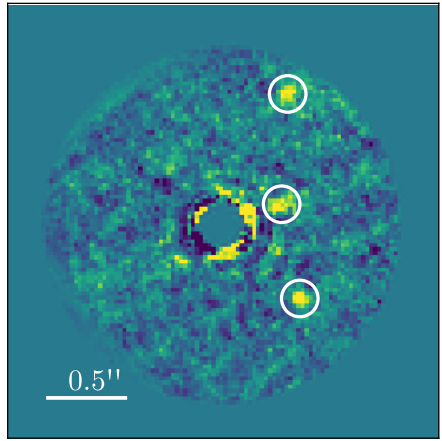}
\end{center}
\caption{\label{fig:8799} Three planets within $\sim1''$ of the star HR~8799
detected with ALES (highlighted with white circles, Stone et al., in prep).
This image shows a sum over the wavelength direction of our ALES spectral cube.
The image of the central star has been removed using principle component
analysis based angular differential imaging.  The inner-most planet is
separated by $0.4''$ and is 9.5 magnitudes fainter than the primary. These data
were collected using a two-point nod sequence.} \end{figure}

Below we show three example ALES datasets
demonstrating our ability to work in the high-contrast environment and to
measure spectra of low-mass companions over a variety of separations. In Figure
\ref{fig:8799} we show the three inner-most directly imaged planets in the
HR~8799 system~\cite{Marois2010}. The image is a sum over the wavelength
dimension since the dataset did not include adequate exposure time to attain
high signal to noise in each spectral slice. The closest-in planet, separated
by $\lesssim0.4''$, is easily separated from the central star, which has
been removed using principle component analysis~\cite{Soummer2012, Amara2012}
on each spectral slice before stacking.  This dataset was collected using the
two-point nod script described above and demonstrates our ability to work in
the high-contrast environment.

\begin{figure} [ht]
\begin{center}
\includegraphics[height=7.5cm]{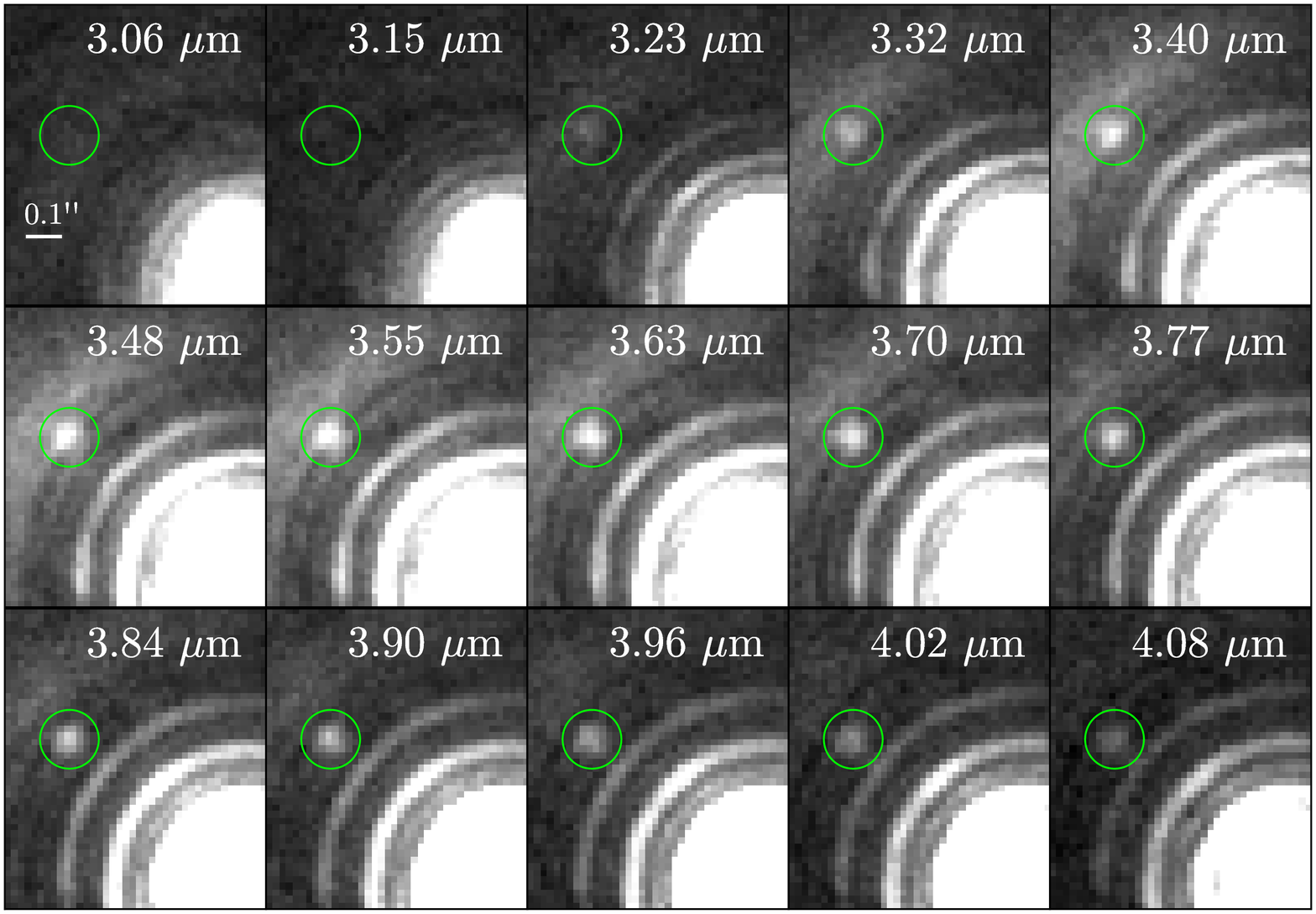}
\end{center}
\caption{ALES spectral cube of the $\kappa$ And system from Stone et al., in
prep. The star is saturated in the lower right corner, and the diffraction
pattern expands with wavelength.  No attempt to remove the stellar image has
been made with these data.  The low-mass companion is highlighted with a green
circle. These data were taken with a three-point nod pattern to center the
companion on the position of the best image quality in the ALES field of view.
\label{fig:kapAnd} } \end{figure}

We also show two datasets obtained using our three-point nod script. In Figure
\ref{fig:kapAnd} we show the $\kappa$~And system that includes
a $\sim20~M_{\mathrm{Jup}}$ companion orbiting a B-type star separated by
$1.1''$~\cite{Carson2013}. We have not made any attempt to remove the image
of the star via post-processing, yet the companion is easily visible at high
signal-to-noise in each spectral slice owing to the high-strehl and
raw-contrast delivered by the LBTI-AO system~\cite{Bailey2014}. In this case,
both the star and the companion just fit into the ALES field of view. We used
the three-point nod script instead of the two-point nod pattern in order to
position the companion where ALES delivers the best optical quality (and
highest spectral resolution and minimal spaxel crosstalk).

\begin{figure} [ht]
\begin{center}
\includegraphics[height=7.5cm]{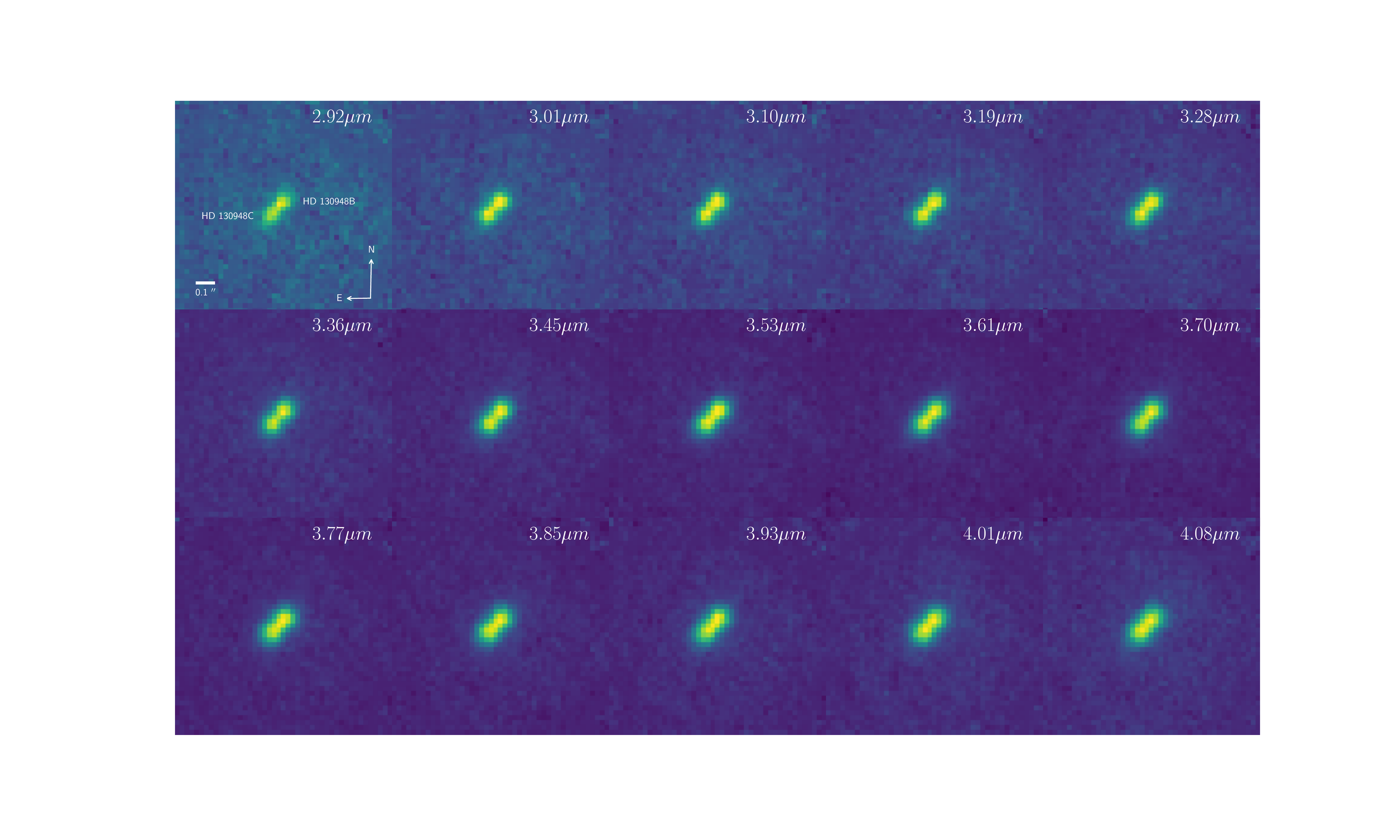}
\end{center}
\caption{ALES spectral cube of HD130948 BC from Briesemeister et al., in prep.
These twin brown dwarfs orbit a main-sequence star as part of a hierarchical
triple system. The star, separated by $2.6''$, was used as the adaptive
optics reference source.  These data were taken using our three-point nod
script. \label{fig:HPBoo}
} \end{figure}

In Figure \ref{fig:HPBoo} we show an ALES data cube of the brown dwarf binary
HD~130948~BC~\cite{Potter2002}. These two brown dwarfs, separated by
$\sim0.1''$, are part of a hierarchical triple system and are in orbit about
a main sequence G star separated by $2.6''$. ALES facilitates resolved
spectroscopy of the brown dwarfs. Since dynamical masses and age constraints
exist for this system, these ALES data can be used to ground truth atmospheric
models for substellar objects at the wavelengths where JWST will operate. For
these observations we used our three-point nod script to observe the brown
dwarfs with ALES while guiding the AO using the G star. 

\acknowledgments 
The authors acknowledge support from NSF Collaborative Research grant 1608834.
JMS is supported by NASA through Hubble Fellow- ship grant HST-HF2-51398.001-A
awarded by the Space Telescope Science Institute, which is operated by the
Association of Universities for Research in Astronomy, Inc., for NASA, under
contract NAS5-26555. The LBT is an international collaboration among
institutions in the United States, Italy and Ger- many. LBT Corporation
partners are: The University of Arizona on behalf of the Arizona university
system; Is- tituto Nazionale di Astrofisica, Italy; LBT Beteiligungs-
gesellschaft, Germany, representing the Max-Planck So- ciety, the Astrophysical
Institute Potsdam, and Heidel- berg University; The Ohio State University, and
The Research Corporation, on behalf of The University of Notre Dame, University
of Minnesota and University of Virginia.
\bibliography{bib} 

\begin{thebibliography}{10}

\bibitem{Kirkpatrick2005}
{Kirkpatrick}, J.~D., ``{New Spectral Types L and T},'' {\em Annual Review of
  Astronomy and Astrophysics}~{\bf 43},  195--245 (Sept. 2005).

\bibitem{Stephens2009}
{Stephens}, D.~C., {Leggett}, S.~K., {Cushing}, M.~C., {Marley}, M.~S.,
  {Saumon}, D., {Geballe}, T.~R., {Golimowski}, D.~A., {Fan}, X., and {Noll},
  K.~S., ``{The 0.8-14.5 {\ensuremath{\mu}}m Spectra of Mid-L to Mid-T Dwarfs:
  Diagnostics of Effective Temperature, Grain Sedimentation, Gas Transport, and
  Surface Gravity},'' {\em ApJ}~{\bf 702},  154--170 (Sept. 2009).

\bibitem{Noll2000}
{Noll}, K.~S., {Geballe}, T.~R., {Leggett}, S.~K., and {Marley}, M.~S., ``{The
  Onset of Methane in L Dwarfs},'' {\em ApJ}~{\bf 541},  L75--L78 (Oct. 2000).

\bibitem{Hinz2016}
{Hinz}, P.~M., {Defr{\`e}re}, D., {Skemer}, A., {Bailey}, V., {Stone}, J.,
  {Spalding}, E., {Vaz}, A., {Pinna}, E., {Puglisi}, A., {Esposito}, S.,
  {Montoya}, M., {Downey}, E., {Leisenring}, J., {Durney}, O., {Hoffmann}, W.,
  {Hill}, J., {Millan-Gabet}, R., {Mennesson}, B., {Danchi}, W., {Morzinski},
  K., {Grenz}, P., {Skrutskie}, M., and {Ertel}, S., ``{Overview of LBTI: a
  multipurpose facility for high spatial resolution observations},'' in [{\em
  Optical and Infrared Interferometry and Imaging
  V}{\nolinebreak\hspace{0.1em}]},   {\bf 9907},  990704 (Aug. 2016).

\bibitem{Skemer2015}
{Skemer}, A.~J., {Hinz}, P., {Montoya}, M., {Skrutskie}, M.~F., {Leisenring},
  J., {Durney}, O., {Woodward}, C.~E., {Wilson}, J., {Nelson}, M., {Bailey},
  V., {Defrere}, D., and {Stone}, J., ``{First light with ALES: A 2-5 micron
  adaptive optics Integral Field Spectrograph for the LBT},'' in [{\em
  Techniques and Instrumentation for Detection of Exoplanets
  VII}{\nolinebreak\hspace{0.1em}]},   {\bf 9605},  96051D (Sept. 2015).

\bibitem{Skrutskie2010}
{Skrutskie}, M.~F., {Jones}, T., {Hinz}, P., {Garnavich}, P., {Wilson}, J.,
  {Nelson}, M., {Solheid}, E., {Durney}, O., {Hoffmann}, W., {Vaitheeswaran},
  V., {McMahon}, T., {Leisenring}, J., and {Wong}, A., ``{The Large Binocular
  Telescope mid-infrared camera (LMIRcam): final design and status},'' in [{\em
  Ground-based and Airborne Instrumentation for Astronomy
  III}{\nolinebreak\hspace{0.1em}]},   {\bf 7735},  77353H (July 2010).

\bibitem{Maire2015}
{Maire}, A.~L., {Skemer}, A.~J., {Hinz}, P.~M., {Desidera}, S., {Esposito}, S.,
  {Gratton}, R., {Marzari}, F., {Skrutskie}, M.~F., {Biller}, B.~A.,
  {Defr{\`e}re}, D., {Bailey}, V.~P., {Leisenring}, J.~M., {Apai}, D.,
  {Bonnefoy}, M., {Brandner}, W., {Buenzli}, E., {Claudi}, R.~U., {Close},
  L.~M., {Crepp}, J.~R., {De Rosa}, R.~J., {Eisner}, J.~A., {Fortney}, J.~J.,
  {Henning}, T., {Hofmann}, K.~H., {Kopytova}, T.~G., {Males}, J.~R., {Mesa},
  D., {Morzinski}, K.~M., {Oza}, A., {Patience}, J., {Pinna}, E., {Rajan}, A.,
  {Schertl}, D., {Schlieder}, J.~E., {Su}, K.~Y.~L., {Vaz}, A., {Ward-Duong},
  K., {Weigelt}, G., and {Woodward}, C.~E., ``{The LEECH Exoplanet Imaging
  Survey. Further constraints on the planet architecture of the HR 8799
  system},'' {\em A\&A}~{\bf 576},  A133 (Apr. 2015).

\bibitem{Hoffmann2014}
{Hoffmann}, W.~F., {Hinz}, P.~M., {Defr{\`e}re}, D., {Leisenring}, J.~M.,
  {Skemer}, A.~J., {Arbo}, P.~A., {Montoya}, M., and {Mennesson}, B.,
  ``{Operation and performance of the mid-infrared camera, NOMIC, on the Large
  Binocular Telescope},'' in [{\em Ground-based and Airborne Instrumentation
  for Astronomy V}{\nolinebreak\hspace{0.1em}]},   {\bf 9147},  91471O (July
  2014).

\bibitem{Brandt2017}
{Brandt}, T.~D., {Rizzo}, M., {Groff}, T., {Chilcote}, J., {Greco}, J.~P.,
  {Kasdin}, N.~J., {Limbach}, M.~A., {Galvin}, M., {Loomis}, C., {Knapp}, G.,
  {McElwain}, M.~W., {Jovanovic}, N., {Currie}, T., {Mede}, K., {Tamura}, M.,
  {Takato}, N., and {Hayashi}, M., ``{Data reduction pipeline for the CHARIS
  integral-field spectrograph I: detector readout calibration and data cube
  extraction},'' {\em Journal of Astronomical Telescopes, Instruments, and
  Systems}~{\bf 3},  048002 (Oct. 2017).

\bibitem{Marois2010}
{Marois}, C., {Zuckerman}, B., {Konopacky}, Q.~M., {Macintosh}, B., and
  {Barman}, T., ``{Images of a fourth planet orbiting HR 8799},'' {\em
  Nature}~{\bf 468},  1080--1083 (Dec. 2010).

\bibitem{Soummer2012}
{Soummer}, R., {Pueyo}, L., and {Larkin}, J., ``{Detection and Characterization
  of Exoplanets and Disks Using Projections on Karhunen-Lo{\`e}ve
  Eigenimages},'' {\em ApJL}~{\bf 755},  L28 (Aug. 2012).

\bibitem{Amara2012}
{Amara}, A. and {Quanz}, S.~P., ``{PYNPOINT: an image processing package for
  finding exoplanets},'' {\em MNRAS}~{\bf 427},  948--955 (Dec. 2012).

\bibitem{Carson2013}
{Carson}, J., {Thalmann}, C., {Janson}, M., {Kozakis}, T., {Bonnefoy}, M.,
  {Biller}, B., {Schlieder}, J., {Currie}, T., {McElwain}, M., {Goto}, M.,
  {Henning}, T., {Brandner}, W., {Feldt}, M., {Kandori}, R., {Kuzuhara}, M.,
  {Stevens}, L., {Wong}, P., {Gainey}, K., {Fukagawa}, M., {Kuwada}, Y.,
  {Brandt}, T., {Kwon}, J., {Abe}, L., {Egner}, S., {Grady}, C., {Guyon}, O.,
  {Hashimoto}, J., {Hayano}, Y., {Hayashi}, M., {Hayashi}, S., {Hodapp}, K.,
  {Ishii}, M., {Iye}, M., {Knapp}, G., {Kudo}, T., {Kusakabe}, N., {Matsuo},
  T., {Miyama}, S., {Morino}, J., {Moro-Martin}, A., {Nishimura}, T., {Pyo},
  T., {Serabyn}, E., {Suto}, H., {Suzuki}, R., {Takami}, M., {Takato}, N.,
  {Terada}, H., {Tomono}, D., {Turner}, E., {Watanabe}, M., {Wisniewski}, J.,
  {Yamada}, T., {Takami}, H., {Usuda}, T., and {Tamura}, M., ``{Direct Imaging
  Discovery of a ''Super-Jupiter'' around the Late B-type Star
  {\ensuremath{\kappa}} And},'' {\em ApJ}~{\bf 763},  L32 (Feb. 2013).

\bibitem{Bailey2014}
{Bailey}, V.~P., {Hinz}, P.~M., {Puglisi}, A.~T., {Esposito}, S.,
  {Vaitheeswaran}, V., {Skemer}, A.~J., {Defr{\`e}re}, D., {Vaz}, A., and
  {Leisenring}, J.~M., ``{Large binocular telescope interferometer adaptive
  optics: on-sky performance and lessons learned},'' in [{\em Adaptive Optics
  Systems IV}{\nolinebreak\hspace{0.1em}]},   {\bf 9148},  914803 (July 2014).

\bibitem{Potter2002}
{Potter}, D., {Mart{\'\i}n}, E.~L., {Cushing}, M.~C., {Baudoz}, P., {Brandner},
  W., {Guyon}, O., and {Neuh{\"a}user}, R., ``{Hokupa'a-Gemini Discovery of Two
  Ultracool Companions to the Young Star HD 130948},'' {\em ApJ}~{\bf 567},
  L133--L136 (Mar. 2002).

\end{thebibliography}
\bibliographystyle{spiebib} 

\end{document}